\documentclass[zpreprint,zbstpl,british]{./zeus_K2.9paper}
\usepackage{./zeusdet_units}
\usepackage{./lineno}
\usepackage{bm}
\usepackage{url}
\chardef\usc=95
\chardef\til=126
\catcode`\@=11 
\DeclareRobustCommand\xdotspace{\futurelet\@let@token\@xdotspace}
\def\@xdotspace{%
  \ifx\@let@token.\else
  \ifx\@let@token\bgroup.\else
  \ifx\@let@token\egroup.\else
  \ifx\@let@token\/.\else
  \ifx\@let@token\ .\else
  \ifx\@let@token~.\else
  \ifx\@let@token!.\else
  \ifx\@let@token,.\else
  \ifx\@let@token:.\else
  \ifx\@let@token;.\else
  \ifx\@let@token?.\else
  \ifx\@let@token/.\else
  \ifx\@let@token'.\else
  \ifx\@let@token).\else
  \ifx\@let@token-.\else
  \ifx\@let@token\@xobeysp.\else
  \ifx\@let@token\space.\else
  \ifx\@let@token\@sptoken.\else
   .\space
   \fi\fi\fi\fi\fi\fi\fi\fi\fi\fi\fi\fi\fi\fi\fi\fi\fi\fi}
\catcode`\@=12 

\newcommand{\stru}[2]{%
   \relax\ifmmode\hbox{\vrule height#1 depth#2 width0pt}%
   \else\vrule height#1 depth#2 width0pt\fi}

\newcommand{\Ronum}[1]{\uppercase\expandafter{\romannumeral#1}}
\newcommand{\ronum}[1]{\expandafter{\romannumeral#1}}
\DeclareRobustCommand{\LaTeXZ}{%
  \LaTeX\kern-.05em4\kern-.1em
  {\raisebox{-0.2ex}{$\scriptstyle\text{ZEUS}$}}\xspace}

\newcommand{\fig}[1]{Fig.~\ref{fig-#1}}
\newcommand{\Fig}[1]{Figure~\ref{fig-#1}}

\newcommand{\Tab}[1]{Table~\ref{tab-#1}}

\newcommand{\sect}[1]{Sect.~\ref{sec-#1}}

\newcommand{\slashfrac}[2]{%
  \raisebox{0.5ex}{\ensuremath #1}\kern-0.12em/\kern-0.08em
  \raisebox{-.8ex}{\ensuremath #2}}

\newcommand{\sqr}[3]{%
    {\vcenter{\hrule height.#3ex\hbox{\vrule width.#2ex height#1ex
     \kern#1ex\vrule width.#3ex}\hrule height.#2ex}}}

\newcommand{\widebar}[1]{%
   \mkern1.5mu\overline{\mkern-1.5mu#1\mkern-1.mu}\mkern1.mu}
\catcode`\@=11 
\newcommand{\parenbar}{\mathpalette\p@renb@r}
\def\p@renb@r#1#2{\vbox{%
  \ifx#1\scriptscriptstyle \dimen@.7em\dimen@ii.2em\else
  \ifx#1\scriptstyle \dimen@.8em\dimen@ii.25em\else
  \dimen@1em\dimen@ii.4em\fi\fi \offinterlineskip
  \ialign{\hfill##\hfill\cr
    \vbox{\hrule width\dimen@ii}\cr
    \noalign{\vskip-.3ex}%
    \hbox to\dimen@{$\mathchar300\hfil\mathchar301$}\cr
    \noalign{\vskip-.3ex}%
    $#1#2$\cr}}}
\catcode`\@=12 

\newcommand{\pbar}{\widebar{p}}

\newcommand{\ubar}{\widebar{u}}

\newcommand{\cbar}{\widebar{c}}
\ifzeusunit
  
\else
  
\fi
\newcommand{\tbar}{\widebar{t}}

\newcommand{\ctwb}{\cos\theta_W}

\newcommand{\tw}{\theta_W}

\newcommand{\IP}{{\rm I$\kern-0.01667em$P}\xspace}

\newcommand{\data}{{\rm data}}

\newcommand{\had}{{\rm had}}
\newcommand{\jet}{{\rm jet}}

\newcommand{\miss}{{\rm miss}}

\mathchardef\qsm=63
\mathchardef\pls=43
\mathchardef\mns=512
\mathchardef\plm=518
\mathchardef\eql=61
\mathchardef\smallleft=300
\mathchardef\smallright=301
\mathchardef\les=316
\mathchardef\gre=318
\mathchardef\leq=532
\mathchardef\grq=533

\catcode`\@=11 
\newcounter{pict@width}
\newcounter{pict@height}
\newlength{\pict@scale}
\newcommand{\psfigadd}[4]{%
\setcounter{pict@width}{1*\ratio{#2+\pict@scale/2}{\pict@scale}}
\setcounter{pict@height}{1*\ratio{#3+\pict@scale/2}{\pict@scale}}
\setlength{\unitlength}{\pict@scale}
\hbox to #2{\hspace{-\fill}\begin{picture}(\thepict@width,\thepict@height)
\put(0,0){\psfig{figure=#1,width=#2,height=#3,clip=}}
\SetScale{0.283466457}
\SetWidth{1.763889}
{#4}
\end{picture}}
}
\newcounter{pict@widthfst}
\newcounter{pict@widthscd}
\newcounter{pict@widthtot}
\newcommand{\psfigaddtwo}[7]{%
\setcounter{pict@widthfst}{1*\ratio{#2+\pict@scale/2}{\pict@scale}}
\setcounter{pict@widthscd}{1*\ratio{#2+#4+\pict@scale/2}{\pict@scale}}
\setcounter{pict@widthtot}{1*\ratio{#2+#4+#6+\pict@scale/2}{\pict@scale}}
\setcounter{pict@height}{1*\ratio{#3+\pict@scale/2}{\pict@scale}}
\setlength{\unitlength}{\pict@scale}
\hbox{\hspace{-\fill}\begin{picture}(\thepict@widthtot,\thepict@height)
\put(0,0){\psfig{figure=#1,width=#2,height=#3,clip=}}
\put(\thepict@widthscd,0){\psfig{figure=#5,width=#6,height=#3,clip=}}
\SetScale{0.283466457}
\SetWidth{1.763889}
{#7}
\end{picture}}
}
\newcommand{\psfigror}[4]{%
\setcounter{pict@width}{1*\ratio{#2+\pict@scale/2}{\pict@scale}}
\setcounter{pict@height}{1*\ratio{#3+\pict@scale/2}{\pict@scale}}
\setlength{\unitlength}{\pict@scale}
\hbox{\begin{picture}(\thepict@width,\thepict@height)
\put(0,\thepict@height){\psfig{figure=#1,width=#3,height=#2,clip=,angle=270}}
\SetScale{0.283466457}
\SetWidth{1.763889}
{#4}
\end{picture}}
}
\newcommand{\psfigrol}[4]{%
\setcounter{pict@width}{1*\ratio{#2+\pict@scale/2}{\pict@scale}}
\setcounter{pict@height}{1*\ratio{#3+\pict@scale/2}{\pict@scale}}
\setlength{\unitlength}{\pict@scale}
\hbox{\begin{picture}(\thepict@width,\thepict@height)
\put(0,0){\psfig{figure=#1,width=#3,height=#2,clip=,angle=90}}
\SetScale{0.283466457}
\SetWidth{1.763889}
{#4}
\end{picture}}
}
\catcode`\@=12 
\newlength\listtextwidth

\catcode`\@=11 
\newlength{\@tabfninsert}
\newlength{\@tabfnwidth}
\newcommand{\tabfootnote}[2]{%
  \setlength{\@tabfninsert}{0.8em}
  \setlength{\@tabfnwidth}{\textwidth}
  \addtolength{\@tabfnwidth}{-\@tabfninsert}
  \addtolength{\@tabfnwidth}{-0.4em}
  \noindent\makebox[\@tabfninsert][r]{\footnotesize$^{#1}$\hfil}\hfill%
  \parbox[t]{\@tabfnwidth}{\footnotesize #2\hfill}}
\catcode`\@=12 

\def\citesingletopCC{{\cite{%
np:b299:21,*np:b304:451,*zp:c51:477}}\xspace}
\def\citeStelzerMoretti{{\cite{%
pr:d56:5919,*pr:d57:3040}}\xspace}
\def\citeGIM{{\cite{%
pr:d2:1285}}\xspace}
\def\citeCKM{{\cite{%
pl:b268:106,*pr:d44:1473,*pl:b307:387}}\xspace}
\def\citeFCNCtreelvl{{\cite{%
pr:d53:1199,*np:d504:45,*np:b337:269,*pl:b457:186}}\xspace}
\def\citesingletopBSM{{\cite{%
pl:b224:423,*np:b337:269,*np:b454:527,*np:d504:45,*pl:b457:186}}\xspace}

\def\citeZEUSsingletop{{\cite{%
pl:b559:153}}\xspace}
\def\citeFCNClagrangian{{\cite{%
pr:d60:074015}}\xspace}

\def\citeCTD{{\cite{%
nim:a279:290,*npps:b32:181,*nim:a338:254%
}}\xspace}
\def\citeMVD{{\cite{%
nim:a581:656%
}}\xspace}
\def\citeSTT{{\cite{%
nim:a535:191%
}}\xspace}
\def\citeCAL{{\cite{%
nim:a309:77,*nim:a309:101,*nim:a321:356,*nim:a336:23%
}}\xspace}
\def\citeHES{{\cite{%
nim:a277:176%
}}\xspace}
\def\citeSixmT{{\cite{%
thesis:gosau:2007%
}}\xspace}
\def\citeBAC{{\cite{%
nim:a300:480%
}}\xspace}
\def\citePCAL{{\cite{%
desy-92-066,*zfp:c63:391,*acpp:b32:2025%
}}\xspace}
\def\citeSPECTRO{{\cite{%
nim:a565:572%
}}\xspace}

\begin{document}
\pagestyle{empty}
%
%
\prepnum{}
\prepdate{}

\title{%
Search for single-top production in $\bm{ep}$ collisions at HERA}
                    
\author{ZEUS Collaboration}
\date{}

\maketitle

%
%
\begin{abstract}\noindent
    A search for single-top production, $ep \rightarrow etX$, has been performed with the
    ZEUS detector at HERA using data corresponding to an integrated luminosity 
    of $0.37\fbi$.
    No evidence for top production was found, consistent with the 
    expecta\-tion from the Standard Model. 
    Limits were computed for single-top production via flavour changing neutral current
    transitions involving a neutral electroweak vector boson, $\gamma$ or $Z$.
\noindent    The result was combined with a previous ZEUS result yielding a total luminosity of $0.50\fbi$.
    A $95\%$ credibility level upper limit of $0.13\pb$ was obtained for the cross section at the
    centre-of-mass energy of $\sqrt{s}=315\gev$.
\end{abstract}
\vspace*{5.0cm}
\begin{center}
{\bf Published in Physics Letters B 708 (2012) 27-36}
\end{center}
\thispagestyle{empty}

\cleardoublepage
\pagestyle{plain}

%
%
%
%


\topmargin-2.0cm
\evensidemargin-0.3cm
\oddsidemargin-0.3cm
\textwidth 16.cm
\textheight 660pt
\parindent0.cm
\parskip0.3cm plus0.05cm minus0.05cm
\def\3{\ss}
\newcommand{\address}{ }
\pagenumbering{Roman}
                                                   %
\begin{center}
{                      \Large  The ZEUS Collaboration              }
\end{center}

{\small


        {\raggedright
H.~Abramowicz$^{45, ah}$, 
I.~Abt$^{35}$, 
L.~Adamczyk$^{13}$, 
M.~Adamus$^{54}$, 
R.~Aggarwal$^{7, c}$, 
S.~Antonelli$^{4}$, 
P.~Antonioli$^{3}$, 
A.~Antonov$^{33}$, 
M.~Arneodo$^{50}$, 
V.~Aushev$^{26, 27, z}$
% ...
        }

\newpage


\makebox[3em]{$^{1}$}
\begin{minipage}[t]{14cm}
{\it Argonne National Laboratory, Argonne, Illinois 60439-4815, USA}~$^{A}$

\end{minipage}\\
\makebox[3em]{$^{2}$}
\begin{minipage}[t]{14cm}
{\it Andrews University, Berrien Springs, Michigan 49104-0380, USA}

\end{minipage}\\
\makebox[3em]{$^{3}$}
\begin{minipage}[t]{14cm}
{\it INFN Bologna, Bologna, Italy}~$^{B}$

\end{minipage}\\
\makebox[3em]{$^{4}$}
\begin{minipage}[t]{14cm}
{\it University and INFN Bologna, Bologna, Italy}~$^{B}$

\end{minipage}\\
\makebox[3em]{$^{5}$}
\begin{minipage}[t]{14cm}
{\it Physikalisches Institut der Universit\"at Bonn,
Bonn, Germany}~$^{C}$

\end{minipage}\\
\makebox[3em]{$^{6}$}
\begin{minipage}[t]{14cm}
{\it H.H.~Wills Physics Laboratory, University of Bristol,
Bristol, United Kingdom}~$^{D}$

\end{minipage}\\
\makebox[3em]{$^{7}$}
\begin{minipage}[t]{14cm}
{\it Panjab University, Department of Physics, Chandigarh, India}

\end{minipage}\\
\makebox[3em]{$^{8}$}
\begin{minipage}[t]{14cm}
{\it Calabria University,
Physics Department and INFN, Cosenza, Italy}~$^{B}$

\end{minipage}\\
\makebox[3em]{$^{9}$}
\begin{minipage}[t]{14cm}
{\it Institute for Universe and Elementary Particles, Chonnam National University,\\
Kwangju, South Korea}

\end{minipage}\\
\makebox[3em]{$^{10}$}
\begin{minipage}[t]{14cm}
{\it Jabatan Fizik, Universiti Malaya, 50603 Kuala Lumpur, Malaysia}~$^{E}$

\end{minipage}\\
\makebox[3em]{$^{11}$}
\begin{minipage}[t]{14cm}
{\it Nevis Laboratories, Columbia University, Irvington on Hudson,
New York 10027, USA}~$^{F}$

\end{minipage}\\
\makebox[3em]{$^{12}$}
\begin{minipage}[t]{14cm}
{\it The Henryk Niewodniczanski Institute of Nuclear Physics, Polish Academy of \\
Sciences, Krakow, Poland}~$^{G}$

\end{minipage}\\
\makebox[3em]{$^{13}$}
\begin{minipage}[t]{14cm}
{\it AGH-University of Science and Technology, Faculty of Physics and Applied Computer
Science, Krakow, Poland}~$^{H}$

\end{minipage}\\
\makebox[3em]{$^{14}$}
\begin{minipage}[t]{14cm}
{\it Department of Physics, Jagellonian University, Cracow, Poland}

\end{minipage}\\
\makebox[3em]{$^{15}$}
\begin{minipage}[t]{14cm}
{\it Deutsches Elektronen-Synchrotron DESY, Hamburg, Germany}

\end{minipage}\\
\makebox[3em]{$^{16}$}
\begin{minipage}[t]{14cm}
{\it Deutsches Elektronen-Synchrotron DESY, Zeuthen, Germany}

\end{minipage}\\
\makebox[3em]{$^{17}$}
\begin{minipage}[t]{14cm}
{\it INFN Florence, Florence, Italy}~$^{B}$

\end{minipage}\\
\makebox[3em]{$^{18}$}
\begin{minipage}[t]{14cm}
{\it University and INFN Florence, Florence, Italy}~$^{B}$

\end{minipage}\\
\makebox[3em]{$^{19}$}
\begin{minipage}[t]{14cm}
{\it Fakult\"at f\"ur Physik der Universit\"at Freiburg i.Br.,
Freiburg i.Br., Germany}

\end{minipage}\\
\makebox[3em]{$^{20}$}
\begin{minipage}[t]{14cm}
{\it School of Physics and Astronomy, University of Glasgow,
Glasgow, United Kingdom}~$^{D}$

\end{minipage}\\
\makebox[3em]{$^{21}$}
\begin{minipage}[t]{14cm}
{\it Department of Engineering in Management and Finance, Univ. of
the Aegean, Chios, Greece}

\end{minipage}\\
\makebox[3em]{$^{22}$}
\begin{minipage}[t]{14cm}
{\it Hamburg University, Institute of Experimental Physics, Hamburg,
Germany}~$^{I}$

\end{minipage}\\
\makebox[3em]{$^{23}$}
\begin{minipage}[t]{14cm}
{\it Imperial College London, High Energy Nuclear Physics Group,
London, United Kingdom}~$^{D}$

\end{minipage}\\
\makebox[3em]{$^{24}$}
\begin{minipage}[t]{14cm}
{\it Institute of Particle and Nuclear Studies, KEK,
Tsukuba, Japan}~$^{J}$

\end{minipage}\\
\makebox[3em]{$^{25}$}
\begin{minipage}[t]{14cm}
{\it Institute of Physics and Technology of Ministry of Education and
Science of Kazakhstan, Almaty, Kazakhstan}

\end{minipage}\\
\makebox[3em]{$^{26}$}
\begin{minipage}[t]{14cm}
{\it Institute for Nuclear Research, National Academy of Sciences, Kyiv, Ukraine}

\end{minipage}\\
\makebox[3em]{$^{27}$}
\begin{minipage}[t]{14cm}
{\it Department of Nuclear Physics, National Taras Shevchenko University of Kyiv, Kyiv, Ukraine}

\end{minipage}\\
\makebox[3em]{$^{28}$}
\begin{minipage}[t]{14cm}
{\it Kyungpook National University, Center for High Energy Physics, Daegu,
South Korea}~$^{K}$

\end{minipage}\\
\makebox[3em]{$^{29}$}
\begin{minipage}[t]{14cm}
{\it Institut de Physique Nucl\'{e}aire, Universit\'{e} Catholique de Louvain, Louvain-la-Neuve,\\
Belgium}~$^{L}$

\end{minipage}\\
\makebox[3em]{$^{30}$}
\begin{minipage}[t]{14cm}
{\it Departamento de F\'{\i}sica Te\'orica, Universidad Aut\'onoma
de Madrid, Madrid, Spain}~$^{M}$

\end{minipage}\\
\makebox[3em]{$^{31}$}
\begin{minipage}[t]{14cm}
{\it Department of Physics, McGill University,
Montr\'eal, Qu\'ebec, Canada H3A 2T8}~$^{N}$

\end{minipage}\\
\makebox[3em]{$^{32}$}
\begin{minipage}[t]{14cm}
{\it Meiji Gakuin University, Faculty of General Education,
Yokohama, Japan}~$^{J}$

\end{minipage}\\
\makebox[3em]{$^{33}$}
\begin{minipage}[t]{14cm}
{\it Moscow Engineering Physics Institute, Moscow, Russia}~$^{O}$

\end{minipage}\\
\makebox[3em]{$^{34}$}
\begin{minipage}[t]{14cm}
{\it Moscow State University, Institute of Nuclear Physics,
Moscow, Russia}~$^{P}$

\end{minipage}\\
\makebox[3em]{$^{35}$}
\begin{minipage}[t]{14cm}
{\it Max-Planck-Institut f\"ur Physik, M\"unchen, Germany}

\end{minipage}\\
\makebox[3em]{$^{36}$}
\begin{minipage}[t]{14cm}
{\it NIKHEF and University of Amsterdam, Amsterdam, Netherlands}~$^{Q}$

\end{minipage}\\
\makebox[3em]{$^{37}$}
\begin{minipage}[t]{14cm}
{\it Physics Department, Ohio State University,
Columbus, Ohio 43210, USA}~$^{A}$

\end{minipage}\\
\makebox[3em]{$^{38}$}
\begin{minipage}[t]{14cm}
{\it Department of Physics, University of Oxford,
Oxford, United Kingdom}~$^{D}$

\end{minipage}\\
\makebox[3em]{$^{39}$}
\begin{minipage}[t]{14cm}
{\it INFN Padova, Padova, Italy}~$^{B}$

\end{minipage}\\
\makebox[3em]{$^{40}$}
\begin{minipage}[t]{14cm}
{\it Dipartimento di Fisica dell' Universit\`a and INFN,
Padova, Italy}~$^{B}$

\end{minipage}\\
\makebox[3em]{$^{41}$}
\begin{minipage}[t]{14cm}
{\it Department of Physics, Pennsylvania State University, University Park,\\
Pennsylvania 16802, USA}~$^{F}$

\end{minipage}\\
\makebox[3em]{$^{42}$}
\begin{minipage}[t]{14cm}
{\it Polytechnic University, Sagamihara, Japan}~$^{J}$

\end{minipage}\\
\makebox[3em]{$^{43}$}
\begin{minipage}[t]{14cm}
{\it Dipartimento di Fisica, Universit\`a 'La Sapienza' and INFN,
Rome, Italy}~$^{B}$

\end{minipage}\\
\makebox[3em]{$^{44}$}
\begin{minipage}[t]{14cm}
{\it Rutherford Appleton Laboratory, Chilton, Didcot, Oxon,
United Kingdom}~$^{D}$

\end{minipage}\\
\makebox[3em]{$^{45}$}
\begin{minipage}[t]{14cm}
{\it Raymond and Beverly Sackler Faculty of Exact Sciences, School of Physics, \\
Tel Aviv University, Tel Aviv, Israel}~$^{R}$

\end{minipage}\\
\makebox[3em]{$^{46}$}
\begin{minipage}[t]{14cm}
{\it Department of Physics, Tokyo Institute of Technology,
Tokyo, Japan}~$^{J}$

\end{minipage}\\
\makebox[3em]{$^{47}$}
\begin{minipage}[t]{14cm}
{\it Department of Physics, University of Tokyo,
Tokyo, Japan}~$^{J}$

\end{minipage}\\
\makebox[3em]{$^{48}$}
\begin{minipage}[t]{14cm}
{\it Tokyo Metropolitan University, Department of Physics,
Tokyo, Japan}~$^{J}$

\end{minipage}\\
\makebox[3em]{$^{49}$}
\begin{minipage}[t]{14cm}
{\it Universit\`a di Torino and INFN, Torino, Italy}~$^{B}$

\end{minipage}\\
\makebox[3em]{$^{50}$}
\begin{minipage}[t]{14cm}
{\it Universit\`a del Piemonte Orientale, Novara, and INFN, Torino,
Italy}~$^{B}$

\end{minipage}\\
\makebox[3em]{$^{51}$}
\begin{minipage}[t]{14cm}
{\it Department of Physics, University of Toronto, Toronto, Ontario,
Canada M5S 1A7}~$^{N}$

\end{minipage}\\
\makebox[3em]{$^{52}$}
\begin{minipage}[t]{14cm}
{\it Physics and Astronomy Department, University College London,
London, United Kingdom}~$^{D}$

\end{minipage}\\
\makebox[3em]{$^{53}$}
\begin{minipage}[t]{14cm}
{\it Faculty of Physics, University of Warsaw, Warsaw, Poland}

\end{minipage}\\
\makebox[3em]{$^{54}$}
\begin{minipage}[t]{14cm}
{\it National Centre for Nuclear Research, Warsaw, Poland}

\end{minipage}\\
\makebox[3em]{$^{55}$}
\begin{minipage}[t]{14cm}
{\it Department of Particle Physics and Astrophysics, Weizmann
Institute, Rehovot, Israel}

\end{minipage}\\
\makebox[3em]{$^{56}$}
\begin{minipage}[t]{14cm}
{\it Department of Physics, University of Wisconsin, Madison,
Wisconsin 53706, USA}~$^{A}$

\end{minipage}\\
\makebox[3em]{$^{57}$}
\begin{minipage}[t]{14cm}
{\it Department of Physics, York University, Ontario, Canada M3J
1P3}~$^{N}$

\end{minipage}\\
\vspace{30em} \pagebreak[4]


\makebox[3ex]{$^{ A}$}
\begin{minipage}[t]{14cm}
 supported by the US Department of Energy\
\end{minipage}\\
\makebox[3ex]{$^{ B}$}
\begin{minipage}[t]{14cm}
 supported by the Italian National Institute for Nuclear Physics (INFN) \
\end{minipage}\\
\makebox[3ex]{$^{ C}$}
\begin{minipage}[t]{14cm}
 supported by the German Federal Ministry for Education and Research (BMBF), under
 contract No. 05 H09PDF\
\end{minipage}\\
\makebox[3ex]{$^{ D}$}
\begin{minipage}[t]{14cm}
 supported by the Science and Technology Facilities Council, UK\
\end{minipage}\\
\makebox[3ex]{$^{ E}$}
\begin{minipage}[t]{14cm}
 supported by an FRGS grant from the Malaysian government\
\end{minipage}\\
\makebox[3ex]{$^{ F}$}
\begin{minipage}[t]{14cm}
 supported by the US National Science Foundation. Any opinion,
 findings and conclusions or recommendations expressed in this material
 are those of the authors and do not necessarily reflect the views of the
 National Science Foundation.\
\end{minipage}\\
\makebox[3ex]{$^{ G}$}
\begin{minipage}[t]{14cm}
 supported by the Polish Ministry of Science and Higher Education as a scientific project No.
 DPN/N188/DESY/2009\
\end{minipage}\\
\makebox[3ex]{$^{ H}$}
\begin{minipage}[t]{14cm}
 supported by the Polish Ministry of Science and Higher Education and its grants
 for Scientific Research\
\end{minipage}\\
\makebox[3ex]{$^{ I}$}
\begin{minipage}[t]{14cm}
 supported by the German Federal Ministry for Education and Research (BMBF), under
 contract No. 05h09GUF, and the SFB 676 of the Deutsche Forschungsgemeinschaft (DFG) \
\end{minipage}\\
\makebox[3ex]{$^{ J}$}
\begin{minipage}[t]{14cm}
 supported by the Japanese Ministry of Education, Culture, Sports, Science and Technology
 (MEXT) and its grants for Scientific Research\
\end{minipage}\\
\makebox[3ex]{$^{ K}$}
\begin{minipage}[t]{14cm}
 supported by the Korean Ministry of Education and Korea Science and Engineering
 Foundation\
\end{minipage}\\
\makebox[3ex]{$^{ L}$}
\begin{minipage}[t]{14cm}
 supported by FNRS and its associated funds (IISN and FRIA) and by an Inter-University
 Attraction Poles Programme subsidised by the Belgian Federal Science Policy Office\
\end{minipage}\\
\makebox[3ex]{$^{ M}$}
\begin{minipage}[t]{14cm}
 supported by the Spanish Ministry of Education and Science through funds provided by
 CICYT\
\end{minipage}\\
\makebox[3ex]{$^{ N}$}
\begin{minipage}[t]{14cm}
 supported by the Natural Sciences and Engineering Research Council of Canada (NSERC) \
\end{minipage}\\
\makebox[3ex]{$^{ O}$}
\begin{minipage}[t]{14cm}
 partially supported by the German Federal Ministry for Education and Research (BMBF)\
\end{minipage}\\
\makebox[3ex]{$^{ P}$}
\begin{minipage}[t]{14cm}
 supported by RF Presidential grant N 4142.2010.2 for Leading Scientific Schools, by the
 Russian Ministry of Education and Science through its grant for Scientific Research on
 High Energy Physics and under contract No.02.740.11.0244 \
\end{minipage}\\
\makebox[3ex]{$^{ Q}$}
\begin{minipage}[t]{14cm}
 supported by the Netherlands Foundation for Research on Matter (FOM)\
\end{minipage}\\
\makebox[3ex]{$^{ R}$}
\begin{minipage}[t]{14cm}
 supported by the Israel Science Foundation\
\end{minipage}\\
\vspace{30em} \pagebreak[4]


\makebox[3ex]{$^{ a}$}
\begin{minipage}[t]{14cm}
now at University of Salerno, Italy\
\end{minipage}\\
\makebox[3ex]{$^{ b}$}
\begin{minipage}[t]{14cm}
now at Queen Mary University of London, United Kingdom\
\end{minipage}\\
\makebox[3ex]{$^{ c}$}
\begin{minipage}[t]{14cm}
also funded by Max Planck Institute for Physics, Munich, Germany\
\end{minipage}\\
\makebox[3ex]{$^{ d}$}
\begin{minipage}[t]{14cm}
also Senior Alexander von Humboldt Research Fellow at Hamburg University,
 Institute of Experimental Physics, Hamburg, Germany\
\end{minipage}\\
\makebox[3ex]{$^{ e}$}
\begin{minipage}[t]{14cm}
also at Cracow University of Technology, Faculty of Physics,
 Mathemathics and Applied Computer Science, Poland\
\end{minipage}\\
\makebox[3ex]{$^{ f}$}
\begin{minipage}[t]{14cm}
supported by the research grant No. 1 P03B 04529 (2005-2008)\
\end{minipage}\\
\makebox[3ex]{$^{ g}$}
\begin{minipage}[t]{14cm}
now at Rockefeller University, New York, NY
 10065, USA\
\end{minipage}\\
\makebox[3ex]{$^{ h}$}
\begin{minipage}[t]{14cm}
now at DESY group FS-CFEL-1\
\end{minipage}\\
\makebox[3ex]{$^{ i}$}
\begin{minipage}[t]{14cm}
now at Institute of High Energy Physics, Beijing, China\
\end{minipage}\\
\makebox[3ex]{$^{ j}$}
\begin{minipage}[t]{14cm}
now at DESY group FEB, Hamburg, Germany\
\end{minipage}\\
\makebox[3ex]{$^{ k}$}
\begin{minipage}[t]{14cm}
also at Moscow State University, Russia\
\end{minipage}\\
\makebox[3ex]{$^{ l}$}
\begin{minipage}[t]{14cm}
now at University of Liverpool, United Kingdom\
\end{minipage}\\
\makebox[3ex]{$^{ m}$}
\begin{minipage}[t]{14cm}
now at CERN, Geneva, Switzerland\
\end{minipage}\\
\makebox[3ex]{$^{ n}$}
\begin{minipage}[t]{14cm}
also affiliated with Universtiy College London, UK\
\end{minipage}\\
\makebox[3ex]{$^{ o}$}
\begin{minipage}[t]{14cm}
now at Goldman Sachs, London, UK\
\end{minipage}\\
\makebox[3ex]{$^{ p}$}
\begin{minipage}[t]{14cm}
also at Institute of Theoretical and Experimental Physics, Moscow, Russia\
\end{minipage}\\
\makebox[3ex]{$^{ q}$}
\begin{minipage}[t]{14cm}
also at FPACS, AGH-UST, Cracow, Poland\
\end{minipage}\\
\makebox[3ex]{$^{ r}$}
\begin{minipage}[t]{14cm}
partially supported by Warsaw University, Poland\
\end{minipage}\\
\makebox[3ex]{$^{ s}$}
\begin{minipage}[t]{14cm}
now at Istituto Nucleare di Fisica Nazionale (INFN), Pisa, Italy\
\end{minipage}\\
\makebox[3ex]{$^{ t}$}
\begin{minipage}[t]{14cm}
now at Haase Energie Technik AG, Neum\"unster, Germany\
\end{minipage}\\
\makebox[3ex]{$^{ u}$}
\begin{minipage}[t]{14cm}
now at Department of Physics, University of Bonn, Germany\
\end{minipage}\\
\makebox[3ex]{$^{ v}$}
\begin{minipage}[t]{14cm}
now at Biodiversit\"at und Klimaforschungszentrum (BiK-F), Frankfurt, Germany\
\end{minipage}\\
\makebox[3ex]{$^{ w}$}
\begin{minipage}[t]{14cm}
also affiliated with DESY, Germany\
\end{minipage}\\
\makebox[3ex]{$^{ x}$}
\begin{minipage}[t]{14cm}
also at University of Tokyo, Japan\
\end{minipage}\\
\makebox[3ex]{$^{ y}$}
\begin{minipage}[t]{14cm}
now at Kobe University, Japan\
\end{minipage}\\
\makebox[3ex]{$^{ z}$}
\begin{minipage}[t]{14cm}
supported by DESY, Germany\
\end{minipage}\\
\makebox[3ex]{$^{\dagger}$}
\begin{minipage}[t]{14cm}
 deceased \
\end{minipage}\\
\makebox[3ex]{$^{aa}$}
\begin{minipage}[t]{14cm}
member of National Technical University of Ukraine, Kyiv Polytechnic Institute,
 Kyiv, Ukraine\
\end{minipage}\\
\makebox[3ex]{$^{ab}$}
\begin{minipage}[t]{14cm}
member of National University of Kyiv - Mohyla Academy, Kyiv, Ukraine\
\end{minipage}\\
\makebox[3ex]{$^{ac}$}
\begin{minipage}[t]{14cm}
partly supported by the Russian Foundation for Basic Research, grant 11-02-91345-DFG\_a\
\end{minipage}\\
\makebox[3ex]{$^{ad}$}
\begin{minipage}[t]{14cm}
Alexander von Humboldt Professor; also at DESY and University of
 Oxford\
\end{minipage}\\
\makebox[3ex]{$^{ae}$}
\begin{minipage}[t]{14cm}
STFC Advanced Fellow\
\end{minipage}\\
\makebox[3ex]{$^{af}$}
\begin{minipage}[t]{14cm}
nee Korcsak-Gorzo\
\end{minipage}\\
\makebox[3ex]{$^{ag}$}
\begin{minipage}[t]{14cm}
This material was based on work supported by the
 National Science Foundation, while working at the Foundation.\
\end{minipage}\\
\makebox[3ex]{$^{ah}$}
\begin{minipage}[t]{14cm}
also at Max Planck Institute for Physics, Munich, Germany, External Scientific Member\
\end{minipage}\\
\makebox[3ex]{$^{ai}$}
\begin{minipage}[t]{14cm}
now at Tokyo Metropolitan University, Japan\
\end{minipage}\\
\makebox[3ex]{$^{aj}$}
\begin{minipage}[t]{14cm}
now at Nihon Institute of Medical Science, Japan\
\end{minipage}\\
\makebox[3ex]{$^{ak}$}
\begin{minipage}[t]{14cm}
now at Osaka University, Osaka, Japan\
\end{minipage}\\
\makebox[3ex]{$^{al}$}
\begin{minipage}[t]{14cm}
also at \L\'{o}d\'{z} University, Poland\
\end{minipage}\\
\makebox[3ex]{$^{am}$}
\begin{minipage}[t]{14cm}
member of \L\'{o}d\'{z} University, Poland\
\end{minipage}\\
\makebox[3ex]{$^{an}$}
\begin{minipage}[t]{14cm}
now at Department of Physics, Stockholm University, Stockholm, Sweden\
\end{minipage}\\
\makebox[3ex]{$^{ao}$}
\begin{minipage}[t]{14cm}
also at Cardinal Stefan Wyszy\'nski University, Warsaw, Poland\
\end{minipage}\\
\pagebreak[4]
}


\cleardoublepage
%
%
%
\pagenumbering{arabic}
\section{Introduction}
\label{sec-int}

     The dominant production process of single top quarks in the Standard Model (SM) 
     in $ep$ collisions\footnote{Here and in the following, $e$ denotes both the
     electron and the positron.} at HERA
     is the charged current (CC) reaction $ep \rightarrow \nu t X$~\citesingletopCC,
     which has a cross section of less than $1 \fb$~\citeStelzerMoretti.
     Flavour changing neutral current (FCNC) processes could enhance single-top production, but they are 
     strongly suppressed in the SM by the GIM mechanism~\citeGIM. This mechanism forbids FCNCs at the 
     tree level, allowing only for small contributions at the one-loop level, exploiting the flavour mixing 
     due to the CKM matrix~\citeCKM.
     Several extensions of the SM predict FCNC contributions already at the tree level~\citeFCNCtreelvl. 
     The search for such new interactions involving the top quark ($ut$ or $ct$ transitions mediated by neutral
     vector bosons, $\gamma$ or $Z$) opens an interesting window to look for effects beyond the SM~\citesingletopBSM. 

The FCNC couplings $tuV$ and $tcV$, with
$V=\gamma,Z$, have been investigated in $p\pbar$ collisions at
the Tevatron, where searches for the 
top-quark decays $t\rightarrow uV$ and $t\rightarrow cV$ \cite{prl:80:2525,
pl:b701:313} were carried out. The Tevatron experiments also constrained the 
couplings $tug$ and $tcg$~\cite{prl:102:151801,*pl:b693:81} which induce FCNC transitions mediated by
the gluon.
The couplings $tuV$ and $tcV$ were also investigated in $e^+e^-$
interactions at LEP2 by searching for single-top production through
the reactions $e^+e^-\rightarrow t\ubar \,(+{\rm c.c.})$ and
$e^+e^-\rightarrow t\cbar \,(+{\rm c.c.})$
\cite{pl:b543:173,pl:b521:181,*pl:b549:290,*pl:b590:21}. 
No evidence for such interactions was found
and limits were set on the branching ratios
\mbox{${\rm Br}(t\rightarrow q\gamma)$} and \mbox{${\rm Br}(t\rightarrow qZ)$},
with $q=u, c$.

\begin{figure}[p]
\begin{center}
\includegraphics[scale=0.70]{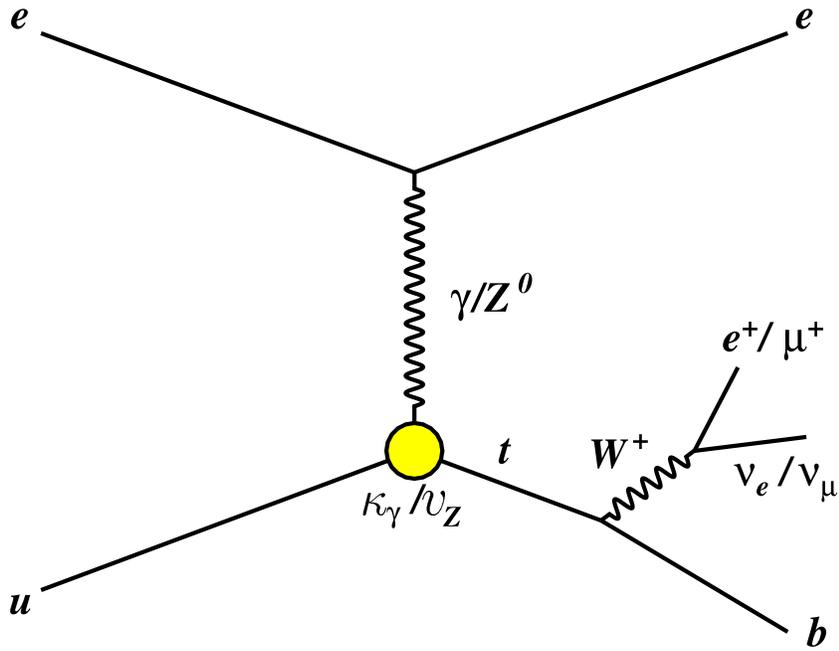}
\end{center}
\caption{
Anomalous single-top production via flavour changing neutral current
 transitions at HERA with subsequent decays $t\rightarrow bW^{+}$ and
$W^{+}\rightarrow \nu_e(\nu_{\mu})e^{+}(\mu^{+})$.}
\label{fig-feynmandiagr}
\end{figure}

The same FCNC couplings could induce single-top production in
$ep$ collisions, 
$ep \rightarrow et X$~\cite{pr:d65:037501}, in which the incoming lepton
exchanges a $\gamma$ or $Z$ with an up quark in the proton, yielding
a top quark in the final state, see \fig{feynmandiagr}. Owing to the large $Z$ mass, this process
is more sensitive to a coupling of the type $tq\gamma$. Furthermore,
large values of $x$, the fraction of the proton momentum carried by the
struck quark, are needed to produce a top quark. Since the $u$-quark
parton distribution function (PDF) of the proton is dominant at large
$x$, the production of single top quark is most sensitive to the 
$tu\gamma$ coupling.

In the present study, the top signal was searched for by looking for the decays 
$t \rightarrow b e \nu_e$ and $t \rightarrow b \mu \nu_{\mu}$. 
At HERA, such event topologies with one lepton with high transverse momentum,
$p_{T}$, and large missing transverse momentum originate predominantly from single-$W$ 
production, which has a cross section of about $1\pb$~\cite{pl:b672:2,*singlewH1,*singlewH1ZEUS}
and is the most important background to any top signal. 
The present analysis extends the previously published ZEUS results~\citeZEUSsingletop which used data 
from the HERA \Ronum{1} running period\footnote{Data collected between 1994 and 2000.}, corresponding to a total 
integrated luminosity of $0.13 \fbi$. The integrated luminosity used in this analysis is about three 
times larger. A combination of the results from the two running periods (total integrated luminosity $0.50 \fbi$) 
has been performed.  


\section{Theoretical framework}
\label{sec-theo}
The effects of the FCNC transitions induced by couplings of the type
$tuV$ are parameterised using the following 
effective Lagrangian~\citeFCNClagrangian:

\begin{equation}
\Delta{\cal L}_{\rm eff} =  e\ e_{t}\ \tbar \frac{i \sigma_{\mu\nu} p^{\nu}}{\Lambda}\ \kappa_{\gamma}\ u\ A^{\mu} + \frac{g}{2\ctwb}\ \tbar \gamma_{\mu}\ v_{Z}\ u Z^{\mu}\ + {\rm h.c.} 
\end{equation}

where $\kappa_{\gamma}$ and $v_{Z}$ are two FCNC couplings mediating $ut$ transitions, $e$ ($e_t$) is the electron (top quark) electric charge,
$g$ is the weak coupling constant, $\tw$ is the weak mixing angle,
$\sigma_{\mu\nu}=\frac{1}{2}(\gamma^{\mu}\gamma^{\nu}-\gamma^{\nu}\gamma^{\mu})$, $\Lambda$ is an effective cut-off parameter which, by convention, is set to the
mass of the $t$ quark, $M_{t}$, $p$ is the momentum of
the gauge boson and $A^{\mu}$ ($Z^{\mu}$) is the photon ($Z$)
field. In the following,
it is assumed that the magnetic coupling $\kappa_{\gamma}$ and the vector coupling
$v_{Z}$ are real and positive. 

The cross section for the process $ep \rightarrow e t X$ was evaluated at the
leading order (LO) using the package CompHEP-4.5.1~\cite{nim:a534:250,*CompHEP} and was parameterised in terms
of three parameters describing the effects of the two FCNC couplings, $A_{\sigma}$ and $B_{\sigma}$,
and their interference, $C_{\sigma}$:
\begin{equation}
\sigma_{ep \rightarrow e t X}=A_{\sigma}\kappa_{\gamma}^{2} + B_{\sigma}v_{Z}^{2} + C_{\sigma}\kappa_{\gamma} v_{Z}.
\end{equation}

The decay widths of the top in the different channels were also evaluated using CompHEP-4.5.1:
\begin{equation}
\Gamma_{t\rightarrow u \gamma}=A_{\Gamma}\kappa_{\gamma}^{2},\ \ \ \
\Gamma_{t\rightarrow u Z}=B_{\Gamma}v_{Z}^{2},\ \ \ \
\Gamma_{t\rightarrow q W}=C_{\Gamma},\ \ \ \
\end{equation}
where $A_{\Gamma}$ and $B_{\Gamma}$ are the partial width of the top corresponding to $u\gamma$ and $uZ$
 unitary FCNC couplings, respectively, and $C_{\Gamma}$ is the SM top width.

The above parameters, summarised in \Tab{pars}, were evaluated using the top mass
$M_{t}=172.0\pm1.6$ \gev~\cite{pdg} and the PDF set CTEQ6L1~\cite{jhep:07:012}.
The interference parameter $C_{\sigma}$ has only a small effect, producing a 
cross section variation of less than $0.5\%$ in the whole range of the couplings considered in this analysis,
and was therefore neglected.
The QCD corrections to the LO cross-section were evaluated at the
approximate next-to-leading order (NLO) and next-to-next-to-leading order
(NNLO)~\cite{pr:d65:037501,jhep:12:04} for magnetic couplings both at 
the $\gamma$ and $Z$ vertices. Since we considered a different 
coupling (vector coupling) at the $Z$ vertex, we used such corrections
only to evaluate the limits for the $\gamma$ exchange (see~\sect{limit1d}). 
Such corrections increase the LO cross-section by $15\%$ and slightly reduces 
the uncertainties due to the QCD factorisation-scale (see~\sect{systematic}).  
The limits involving both coupling (see~\sect{limit2d}) were evaluated using
the LO cross-section.

\section{Experimental setup}
\label{sec-exp}

     The analysis is based on $ep$ collisions recorded with the ZEUS detector during the
     HERA \Ronum{2} running period\footnote{Data collected between 2004 and 2007.}, using
     an integrated luminosity of $0.37 \fbi$, divided into two approximately equal samples of
     $e^+p$ and $e^-p$ collisions.
     The lepton beams were polarised, with roughly equal luminosities
     for positive and negative polarisation, such that the average 
     polarisation was negligible for this analysis.

A detailed description of the ZEUS detector can be found elsewhere~\cite{zeus:1993:bluebook}. 
A brief outline of the components that are most relevant for this analysis is given below.

 Charged particles were tracked in the central tracking detector (CTD)~\cite{nim:a279:290,*npps:b32:181,*nim:a338:254}
 which operated in a magnetic field of $1.43\Tesla$ provided by a thin
 superconducting solenoid. The CTD consisted of 72~cylindrical drift chamber
 layers, organised in nine superlayers covering the
 polar-angle\footnote{The ZEUS coordinate system is a right-handed Cartesian
 system, with the $Z$ axis pointing in the proton beam direction,
 referred to as the ``forward direction'', and the $X$ axis pointing towards  
 the centre of HERA. The coordinate origin is at
 the nominal interaction point. 
 The pseudorapidity is defined as \mbox{$\eta=-\ln\left(\tan\frac{\theta}{2}\right)$},
 where the polar angle, $\theta$, is measured with respect to the proton beam
 direction.} 
 region \mbox{$15^\circ<\theta<164^\circ$}. 
 The CTD was complemented by a silicon microvertex detector
 (MVD)~\cite{nim:a581:656}, consisting of three active layers in the barrel
 and four disks in the forward region.
 For CTD-MVD tracks that pass through all nine CTD superlayers, the momentum 
 resolution was $\sigma(p_{T})/p_{T} = 0.0029p_{T} \oplus 0.0081 \oplus 0.0012/p_{T}$
 with $p_{T}$ in GeV.

The high-resolution uranium--scintillator calorimeter (CAL)~\cite{nim:a309:77,*nim:a309:101,*nim:a321:356,*nim:a336:23}
consisted of three parts: the forward (FCAL), the barrel (BCAL) and the rear (RCAL)
calorimeters. Each part was subdivided transversely into towers and
longitudinally into one electromagnetic section (EMC) and either one (in RCAL)
or two (in BCAL and FCAL) hadronic sections (HAC). The smallest subdivision of
the calorimeter was called a cell.  The CAL energy resolutions, as measured under
test-beam conditions, were $\sigma(E)/E=0.18/\sqrt{E}$ for electrons and
$\sigma(E)/E=0.35/\sqrt{E}$ for hadrons, with $E$ in $\Gev$.


 The luminosity was measured using the Bethe-Heitler reaction 
$ep  \rightarrow e \gamma p$ by a luminosity detector which consisted of 
a lead--scintillator calorimeter~\cite{Desy-92-066,*zfp:c63:391,*acpp:b32:2025}
and an independent magnetic spectrometer~\cite{nim:a565:572}.
The fractional uncertainty on the measured luminosity was $1.9\%$.

\section{Monte Carlo simulation}   

Samples of events were generated using Monte Carlo (MC)
simulations to determine the selection efficiency for 
single-top events produced through FCNC processes and to estimate
background rates from SM processes. The generated events were passed
through the {\sc Geant-3.21}~\cite{tech:cern-dd-ee-84-1} ZEUS
detector- and trigger-simulation programs
\cite{zeus:1993:bluebook}. They were reconstructed and analysed by the
same program chain as the data.

Single-top samples were generated with {\sc Comphep} 4.5.1,
interfaced with {\sc Pythia} 6.14 \cite{cpc:135:238} for  
parton showering, hadronisation and particle decay.
The mass of the top quark in {\sc Comphep} was set to 
     $M_{t} = 175 \gev$. Different sets
     were produced for the two 
     different production processes
     ($\gamma$- and $Z$-mediated) and for the two decay modes 
     ($t \rightarrow bW$ and $t \rightarrow uZ$).
 
     Alternative sets were also generated, only for the $\gamma$-mediated process, with
     the {\sc Hexf} generator~\cite{lsuhe-145-1993} assuming top-quark masses 
     of $170$ and $175$~GeV.
     These sets were used to study the small effect of $M_{t}$ variation, in order to correct the 
     selection efficiency, evaluated using the {\sc Comphep} samples, for the different $M_t$ values 
     used in the generation and in the cross-section calculation (see~\sect{theo}).
     Initial-state radiation from the lepton beam was included using the Weizs\"acker-Williams
     approximation~\cite{prep:146:1}. The hadronic final state was simulated using
     the matrix-element and parton-shower model of 
     {\sc Lepto}~\cite{cpc:101:108}
     for the QCD cascade and the Lund string model~\cite{prep:97:31} as
     implemented in {\sc Jetset}~\cite{cpc:39:347,*cpc:43:367} for the
     hadronisation. 
     The results for {\sc Comphep} and the alternative samples agree within uncertainties.  

     Standard Model single-$W$ production is the most significant background to
     top production.  
     Another important background in the electron-decay channel of the $W$
     ($t\rightarrow bW\rightarrow be\nu$) arises from neutral current (NC)
     deep inelastic scattering (DIS). In addition, two-photon processes provide a 
     source of high-$p_T$ leptons that are a significant background in the
     muon-decay channel of the $W$ ($t\rightarrow bW\rightarrow b\mu\nu$). 
     The CC DIS is a minor source of background for both channels.

     The following MC programs were used to simulate the different background
     processes.
Single-$W$ production was simulated using the event generator 
{\sc Epvec}~\cite{np:b375:3} which did not include hard QCD radiation. 
The $ep \rightarrow eWX$ and $ep \rightarrow \nu WX$ events from 
{\sc Epvec} were scaled by a factor dependent on the transverse momentum and rapidity
of the $W$, such that the resulting cross section corresponded to a 
calculation including QCD corrections at next-to-leading order 
\cite{Diener:2002if,*Nason:1999xs}.

Neutral current and CC DIS events were simulated using the {\sc Lepto}~6.5 
program~\cite{cpc:101:108}, interfaced to 
{\sc Heracles}~4.6.1~\cite{cpc:69:155,*spi:www:heracles}
via {\sc Djangoh}~1.1~\cite{cpc:81:381,*spi:www:djangoh11}. 
The {\sc Heracles} program includes photon
and $Z$ exchanges and first-order electroweak radiative
corrections. The QCD cascade was modelled with the colour-dipole
model~\cite{Azimov:1986sf,*Gustafson:1986db,*np:b306:746,*Andersson:1988gp} by
using the {\sc Ariadne}~4.08 program~\cite{cpc:71:15,*zfp:c65:285}.

Two-photon processes were simulated using the generator
{\sc Grape}~1.1~\cite{cpc:136:126}, which includes dilepton production via
$\gamma\gamma$, $Z\gamma$ and $ZZ$ processes and considers both
elastic and inelastic production at the proton vertex.


    \section{Event selection}   
     
The event selection was optimised for single-top production via photon exchange,
looking for the dominant decay $t \rightarrow bW $ and
subsequent $W$ decay to $e$ and $\mu$ and their respective neutrinos. 
The selection is based on requiring an isolated high-$p_{T}$ lepton and a large missing 
transverse momentum.

     Cosmic background, relevant especially for the muon channel, was suppressed
     using timing cuts based on calorimeter measurements and the track impact 
     parameter with respect to the beam spot. 
     Further cosmic background overlapping with $ep$ interactions was rejected
     by applying a cut $E-p_{Z} < 60 \gev$, $E-p_{Z}$ being the sum 
     of the total and longitudinal energy deposits of the cells in the calorimeter.
     For fully contained events, $E-p_{Z}$ is twice the electron-beam energy and peaks 
     at $55 \gev$.

     Events from beam-gas interactions were rejected on the basis of the ratio of the number
     of tracks pointing to the vertex to the total number of tracks in an event. 

\subsection{Online selection}
\label{sub-trigcon}
A three-level trigger system was used to select events
online~\cite{uproc:chep:1992:222,*nim:a580:1257}.
At the first level, coarse calorimeter and tracking information
were available.
Events were selected using criteria based on either the transverse
energy or missing transverse momentum measured in the CAL.
Events were accepted with a low threshold on these quantities when a
coincidence with CTD tracks from the event vertex was found, while
a higher threshold was used for events with no CTD tracks.

At the second level, timing information from the CAL was used to reject
events inconsistent with an $ep$ interaction. In addition, the
topology of the CAL energy deposits was used to reject non-$ep$ background
events. In particular, a tighter cut was made on missing transverse momentum, since
the resolution in this variable was better at the second than at the first
level.

At the third level, track reconstruction and vertex finding were
performed and used to reject events with a vertex inconsistent with
$ep$ interactions. 
Cuts were applied to calorimeter quantities and reconstructed tracks
to further reduce beam-gas contamination.

     \subsection{Offline selection} 

Jets, used in the selection to define lepton isolation, were reconstructed from CAL cells using the $k_{T}$
cluster algorithm \cite{np:b406:187} in the longitudinally invariant inclusive
mode \cite{pr:d48:3160} and were corrected
for energy loss due to the dead material in front of the CAL. The jets
were required to have a transverse energy $E_{T}^{\jet} > 4.5 \gev$ and
pseudorapidity $|\eta^{\jet}| < 2.5$.

{\bf Muon selection}

      Muons were reconstructed by matching calorimeter cell-patterns compatible with a 
      minimum-ionising particle to CTD tracks \cite{nim:a453:336}. 
      Events were selected as follows: 
      \begin{itemize}
         \item{$\mid Z_{\rm vtx} \mid < 30 \cm$, $Z_{\rm vtx}$ being the $Z$ coordinate of the interaction vertex, 
               to restrict to a region compatible with $ep$ interactions;}
         \item{$ E-p_{Z} > 10 \gev$.
The $E-p_{Z}$ of the CAL deposit associated with
the muon was replaced by that of the muon track. 
This requirement rejected photoproduction events, which populate the low $E-p_{Z}$ region;}
         \item{$P_{T}^{\miss} > 10 \gev$, $P_{T}^{\miss}$ being the missing transverse momentum measured by the CAL;}
         \item{at least one muon candidate with the following characteristics:}
          \begin{itemize}
           \item{a track from the primary vertex matched with a
                 CTD track with at least three hit superlayers 
                 and a transverse momentum, $p_{T}^{\mu}$, greater than $8 \gev$;}
           \item{the distance, $\Delta R$, of the muon candidate in the pseudorapidity-azimuth 
                ($\eta$-$\phi$) plane with respect to any other track and jet in the event satisfying  
                 $\Delta R = \sqrt{(\Delta\eta)^2+(\Delta\phi)^2} > 0.5$.} 
          \end{itemize}
      \end{itemize}

       A total of $269$ events were selected, while $260 \pm 3$ (stat.)
       were expected from the SM, which is dominated by the dimuon production from 
       the $\gamma\gamma$ process. 
       The quoted uncertainty is the error on the expected SM prediction due to the MC 
       statistics.        

\begin{figure}[p]
\vfill
\begin{center}
\includegraphics[scale=0.82]{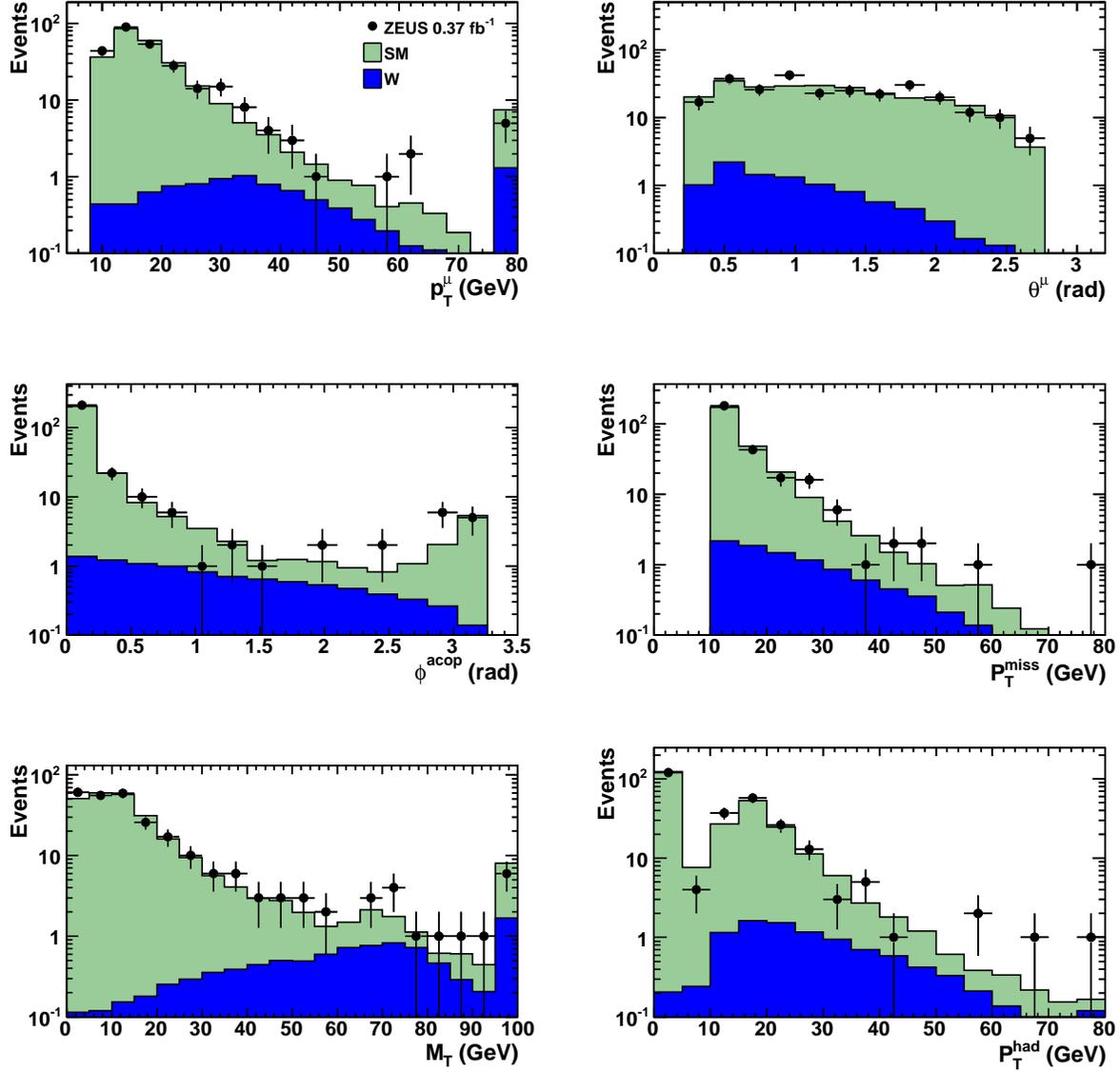}
\end{center}
\caption{Comparison between data and SM expectations for several 
variables in the muon channel:
$p_{T}^{\mu}$, $\theta^{\mu}$, $\phi^{\rm acop}$, $P_{T}^{\miss}$, 
$M_{T}$, $P_{T}^{\had}$.
The contribution of single-$W$ production is also shown as the dark-shaded region.
Any histogram overflows are included in the last bin.}
\label{fig-mu_prel}
\vfill
\end{figure}

      \Fig{mu_prel} shows the comparison between data and
      MC for the variables $p_{T}^{\mu}$, $\theta^{\mu}$, 
      acoplanarity ($\phi^{\rm acop}$), $P_T^{\miss}$,
      transverse mass ($M_T$), 
      hadronic transverse momentum ($P_{T}^{\had}$). Here $P_{T}^{\had}$, $M_T$ and $\phi^{\rm acop}$ are defined
      as follows:
      \begin{itemize}
      \item[-] $P_{T}^{\had} = \sqrt{(\sum_i{P_X^{i}})^2 + (\sum_i{P_Y^{i}})^2}$, where $P_X^{i}$ and 
      $P_Y^{i}$ are the $X$ and $Y$ components of the CAL
      energy deposits not associated with the lepton; 
      \item[-] $M_T = \sqrt{2p_T^{l}p_T^{\nu}(1 - \cos{\phi^{l\nu}})}$, where $p_T^{l}$ is the
      lepton transverse momentum, $p_T^{\nu}$ is the modulus of the missing $P_T$ vector 
      obtained from the CAL and corrected using track information to account for muons,
      $\phi^{l\nu}$ is the azimuthal separation between the lepton and the 
      missing $P_T$ vector; 
      \item[-] $\phi^{\rm acop}$ 
      is the angle
      between the lepton and the vector balancing the $P_{T}^{\rm had}$ and 
      is defined for events with $P_{T}^{\rm had}$ greater than $1 \gev$.
\end{itemize}
      Reasonable agreement is observed in all cases.

{\bf Electron selection}

      Electrons were reconstructed using an algorithm that combined information from 
      the cluster of the energy deposits in the
      calorimeter with tracks \cite{epj:c11:427}. Events were selected as follows:
      
      \begin{itemize}
        \item{$\mid Z_{\rm vtx} \mid < 30 \cm$;}
        \item{$5 < E-p_{Z} < 50 \gev$, to reject NC DIS and photoproduction background;}
        \item{$P_{T}^{\miss} > 12 \gev$;}
        \item{at least one electron candidate with the following characteristics:}
         \begin{itemize}
           \item[-]{$p_{T}^{\rm el} > 10 \gev$;}
           \item[-]{$ 0.3 < \theta^{\rm el} < 2 \rad$;}
           \item[-]{isolated from other tracks and jets 
                    in the event, $\Delta R >0.5$;}
           \item[-]{the extrapolation of the track associated with the electron
                    into the CAL should have a 
                    distance of closest approach to the CAL cluster centre
                    $ < 10  \cm$ and a reconstructed momentum $p > 5 \gev$;}
         \end{itemize}     
        \item{$M_{T} > 10 \gev$, to reject events with 
                        $P_{T}^{\miss}$ along the electron direction;} 
        \item{$0.1 < \phi^{\rm acop} < (\pi - 0.1) \rad$, 
              to reject badly reconstructed NC DIS events with $P_{T}^{\miss}$ in 
              the direction of the electron or of the jet.} 
      \end{itemize}
 
      A total of $245$ events were selected, while $253 \pm 6$ (stat.) 
      were expected from the SM, which is dominated by the NC DIS process.
      The quoted uncertainty is the error on the expected SM prediction due to the MC 
      statistics.        

\begin{figure}[p]
\vfill
\begin{center}
\includegraphics[scale=0.82]{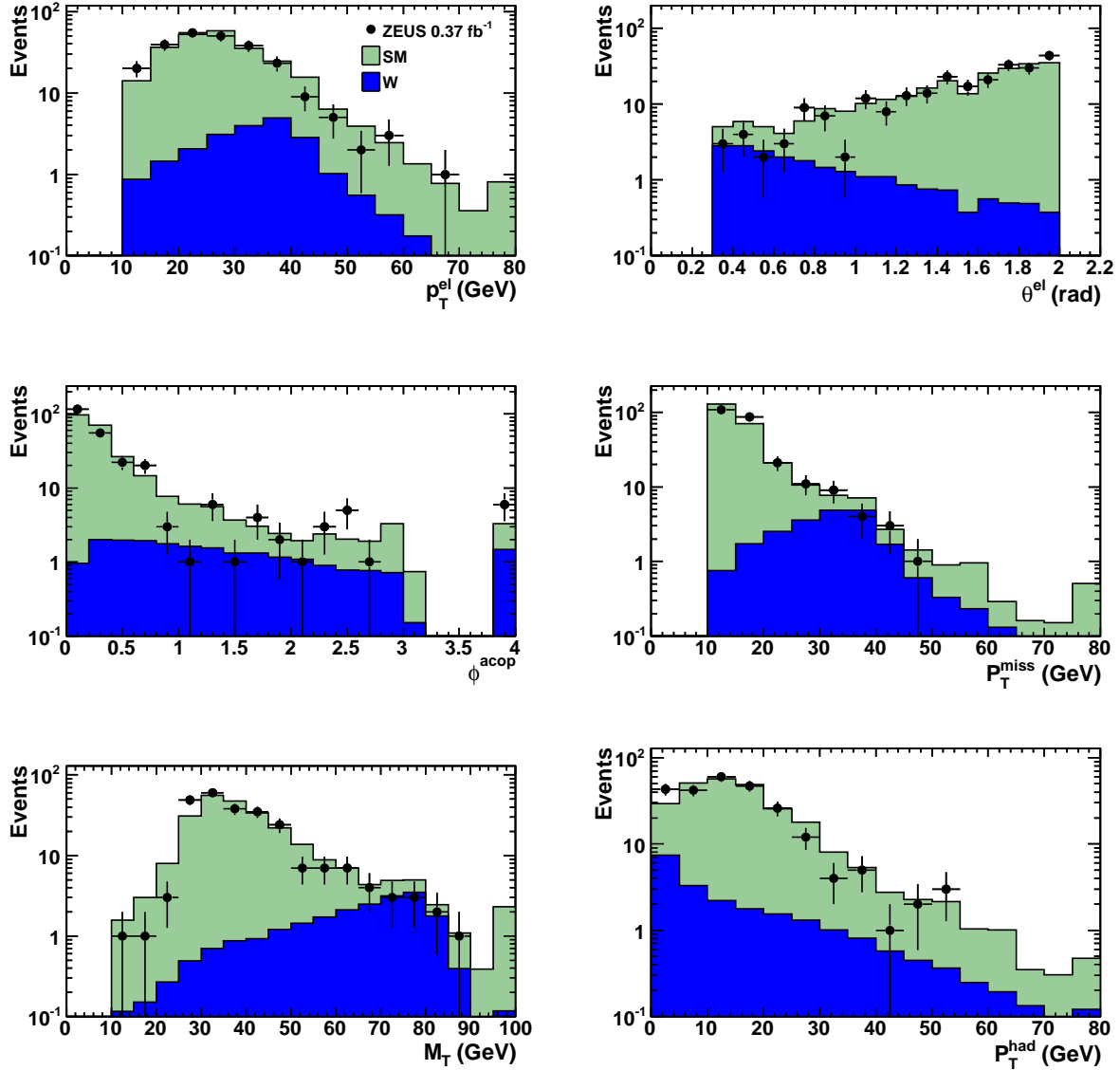}
\end{center}
\caption{Comparison between data and SM expectations for several variables
in the electron channel:
$p_{T}^{\rm el}$, $\theta^{\rm el}$, $\phi^{\rm acop}$, $P_{T}^{\miss}$, 
$M_{T}$, $P_{T}^{\had}$.
The contribution of single-$W$ production is also shown as the dark-shaded region. 
The last bin of the $\phi^{\rm acop}$ histogram
contains events with $P_{T}^{\had}$ less than $1 \gev$
for which $\phi^{\rm acop}$ was not evaluated. In the other cases,
any overflows are included in the last bin.}
\label{fig-el_prel}
\vfill
\end{figure}

      \Fig{el_prel} shows the comparison between data and
      MC for the variables $p_{T}^{\rm el}$, $\theta^{\rm el}$, 
      $\phi^{\rm acop}$, $P_T^{\miss}$, $M_T$, $P_{T}^{\had}$. Reasonable agreement 
      is observed in all cases.

     \subsection{Selection of single-top candidates}
     \label{sec-singletopsel} 

      Since no excess of events above the SM expectation was observed,
      a further selection was made to maximise the sensitivity to a possible
      FCNC single top signal.
      A cut on $P_{T}^{\had}$ of $40 \gev$ was applied to 
      both decay channels while the cuts
      on $\phi^{\rm acop}$ and $P_{T}^{\miss}$ were optimised separately for
      the two channels:
      \begin{itemize}
       \item{$P_{T}^{\had} > 40 \gev$ for both channels;}
       \item[]{\textbf{muon channel:}}
        \begin{itemize}
         \item{$\phi^{\rm acop} > 0.05 \rad$;}
         \item{events with more than one isolated muon were rejected;}
        \end{itemize}
       \item[]{\textbf{electron channel:}}
        \begin{itemize}
         \item{$\phi^{\rm acop} > 0.15 \rad$;}
         \item{$P_{T}^{\miss} > 15 \gev$.}
        \end{itemize}
      \end{itemize}

\begin{figure}[p]
\vfill
\begin{center}
\includegraphics[scale=0.8]{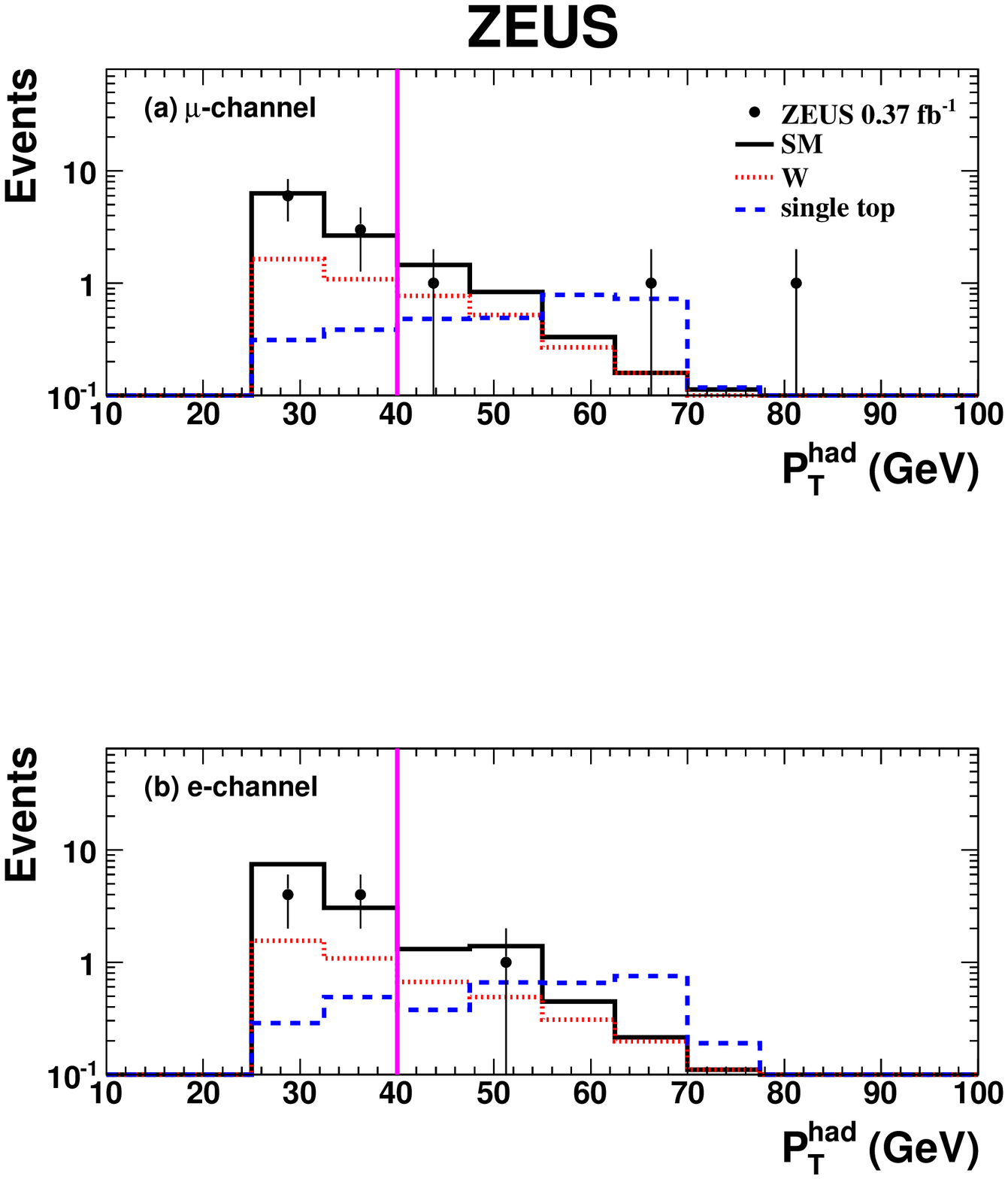}
\end{center}
\caption{Comparison between data and MC expectations for the
         $P_{T}^{\had}$ distribution applying the final selection with a relaxed
         $P_{T}^{\had}$ cut at $25 \gev$ for (a) the muon and (b) the electron channel.
         The dots are the data, the solid histogram is the SM prediction including the 
         $W$ contribution, the dotted histogram the $W$ contribution alone and the 
         dashed histogram the single-top distribution normalized to the limit
         on the signal cross section of $0.24 \pb$ (see~\sect{limit1d}).
         The final selection cut, $P_{T}^{\had} > 40 \gev$, is indicated.}
\label{fig-pthad25}
\vfill
\end{figure}

      One event survived the selection cuts in the electron channel 
      while three events were found in the muon channel.
      \Tab{finalevts} summarises the results of the final selection. 
      In order to compare the MC to data, 
      the $P_{T}^{\had}$ cut was relaxed to $25 \gev$. Figures~\ref{fig-pthad25} (a) and (b) show
      the $P_{T}^{\had}$ behaviour for data and SM expectations
      for the muon and electron channels, respectively. 
      Good agreement between data and predictions is observed for both channels.
      Also shown are the expectations for top production through FCNC, normalised to
      the limit on the signal cross section obtained in~\sect{limit1d}. 
      The data do not support a significant contribution from this process.


\section{Systematic uncertainties}
\label{sec-systematic}

The following systematic uncertainties were taken into account:

\begin{itemize}
\item the theoretical uncertainty on the $W$ background normalisation was assumed to 
      be $\pm 15\%$\cite{Diener:2002if,*Nason:1999xs};
\item the statistical uncertainty on the total SM prediction after the final selection
      was $\pm 13\%$ and $\pm 9\%$ for the $e$- and $\mu$-channel, respectively;
\item the uncertainty on the NC DIS background, particularly relevant for the $e$-channel,
      was evaluated using a sample of events enriched in NC DIS by replacing the $E-p_{Z}$ and 
      acoplanarity cuts by $E-p_{Z}>40 \gev$
      and $\phi^{\rm acop} < 0.3$. A systematic uncertainty of $\pm 15\%$ on this source was determined
      by the level of agreement between data and MC for such a selection. The effect of this 
      uncertainty on the final selection SM prediction was $\pm6\%$ for the $e$-channel and
      negligible for the $\mu$-channel;
\item the uncertainty on the electromagnetic and the hadronic CAL energy scale
      was assumed to be $\pm 1\%$ and $\pm 2\%$, respectively.
      The two scale uncertainties, summed in quadrature, produced a variation 
      of $\pm 6\%$ and of $\pm 5\%$ on the final SM predictions for 
      the $e$- and the $\mu$-channel, respectively, while the effect on the signal selection efficiencies
      was below $2 \%$ and was therefore neglected;
\item the uncertainty on the top mass, $172.0\pm 1.6 \gev$ \cite{pdg}, produced a variation on the
      parameters of the signal cross section and decay widths as reported in \Tab{pars} and a variation 
      of $\pm 2\%$ on the signal selection efficiencies;
\item the uncertainties on the signal efficiency due to the statistics of the MC samples are
reported in \Tab{eff} for the different channels and decay processes;
\item the uncertainties on the PDFs gave a variation on the parameters of the signal cross section 
      as reported in \Tab{pars}. Such uncertainties were evaluated as suggested by the CTEQ 
      group~\cite{jhep:07:012};
\item the uncertainty due to the QCD factorisation-scale affected the signal cross section
      by $\pm 9\%$ for the LO calculation and by $^{+8\%}_{-7\%}$ including the approximated NLO and 
      NNLO QCD corrections (see~\sect{theo}). 
      This effect was evaluated by varying the central value, set to $M_{t}$, between 
      $M_{t}/2$ and $2M_{t}$;
\item the uncertainty on the luminosity determination was $\pm 1.9 \%$.
\end{itemize}

The uncertainties due to the $W$ normalisation, CAL energy scale, top mass,
PDFs and luminosity were assumed to be correlated for the different channels and datasets.
All the above uncertainties were included in the limit calculation as explained in 
\sect{limit1d}.


\section{Limits on FCNC}
\label{sec-limitsonfcnc}
Since no excess over the SM prediction was observed, limits on FCNC couplings
of the type $tuV$ were evaluated using the results of \Tab{finalevts}.
As a first step, limits were evaluated
on the signal cross section and on the $\kappa_{\gamma}$ coupling assuming $v_{Z}=0$.
In a second step, the effect of a non-zero $v_{Z}$ coupling was accounted for.
Limits on the anomalous top branching ratios, {\rm Br}($t \rightarrow u \gamma$) ({\rm Br}$_{u\gamma}$) and
{\rm Br}($t \rightarrow u Z$) ({\rm Br}$_{uZ}$), were evaluated.

\subsection{Limits on the cross section and \bm{$\kappa_{\gamma}$} }
\label{sec-limit1d}

The limit on the anomalous top-production cross section was evaluated using a Bayesian approach and assuming a 
constant prior in the cross section, $\sigma$:

\begin{eqnarray}
f(\sigma|\data) &=& \frac{\prod_i P(N_{i}^{\rm obs}|\sigma)f_0(\sigma)}{\int_0^{\infty} \prod_i P(N_{i}^{\rm obs}|\sigma)f_0(\sigma)d\sigma},\\
P(N_{i}^{\rm obs}|\sigma)&=&\frac{\mu_i^{N_{i}^{\rm obs}}e^{-\mu_i}}{N_{i}^{\rm obs}!},\\ \nonumber
\mu_i&=&N_{i}^{\rm sig}+N_{i}^{\rm bg},\\ \nonumber
N_{i}^{\rm sig}&=&\sigma \mathcal{L}_i \epsilon_i, \\ \nonumber
\label{eq:eq3}
\end{eqnarray}
where $f(\sigma|\data)$ is the posterior probability density function (p.d.f.) of the signal cross section, 
$f_0(\sigma)$ its prior, $i$ runs over the different channels and datasets, $N_{i}^{\rm obs}$ is the number 
of events surviving 
the event selection, $N_{i}^{\rm sig}$ and $N_{i}^{\rm bg}$ are the number of 
signal events and the expected SM background, $\mathcal{L}_i$ is the 
integrated luminosity and $\epsilon_i$ the signal efficiency including branching ratio for each decay channel
(see the first row in \Tab{eff}). The branching ratio of the top to $u\gamma$ was taken into account in the limits
evaluation, the selection efficiency for such channel is expected to be low and was therefore set to zero. 
The systematic uncertainties were treated as nuisance parameters (NPs) and included in the limit calculation, integrating out their dependence (marginalisation) assuming Gaussian priors\footnote{In case of unphysical values, the Gaussian 
priors were truncated.}.
The marginalisation over the NPs and the extraction of the posterior p.d.f. was performed using the package {\sl Bayesian Analysis Toolkit} \cite{bat}, which carries out multidimensional integration using the Markov Chain Monte Carlo technique.

       The $95\%$ Credibility Level (C.L.) limit on the cross section 
       was evaluated by integrating the posterior p.d.f.
       \begin{equation}
        \centering
        \int_{0}^{\sigma_{95}}f(\sigma|\data)d\sigma = 0.95,
       \end{equation}
and found to be
       \begin{equation}
        \centering
        \sigma < 0.24 \pb\ (95\%~{\rm C.L.})~{\rm at}~\sqrt{s}=318\gev. 
       \end{equation}
The limit on the cross section was converted into a limit on the coupling
$\kappa_{\gamma}$, assuming a vanishing $v_{Z}$ coupling and using the $A_{\sigma}$ 
parameter described in \sect{theo} taking into account the approximated NLO and NNLO
QCD corrections (see~\sect{theo}):
       \begin{equation}
        \centering
         \kappa_{\gamma}< 0.17\ (95\%~{\rm C.L.}).
       \end{equation}

      The limit is similar to that obtained by ZEUS from HERA \Ronum{1} data
      \cite{singletophera1} with an integrated luminosity of $0.13 \fbi$.
      In the HERA \Ronum{1} data, no events were found in either the electron 
      or muon channel and also the hadronic $W$-decay channel was exploited.\\

      \vspace{0.1cm}

      The present result was combined with the HERA \Ronum{1} limit for a total 
integrated luminosity of $0.50 \fbi$, using the same Bayesian approach 
as described above and assuming full correlation for the systematic uncertainties due to
the $W$ normalisation, CAL energy scale, top mass and PDFs.
 
      The combined cross-section and $\kappa_{\gamma}$ limits are:
       \begin{equation}
        \centering
        \sigma < 0.13 \pb\ (95\%~{\rm C.L.})~{\rm at}~\sqrt{s}=315\gev,
       \end{equation}
       \begin{equation}
        \centering
          \kappa_{\gamma} < 0.12\ (95\%~{\rm C.L.}).
       \end{equation}
 
      \vspace{0.1cm}
      The combined cross-section limit corresponds to a centre-of-mass energy of $315 \gev$
      since part of the HERA \Ronum{1} data was collected at $\sqrt{s}=300\gev$.


\subsection{Limits on the top anomalous branching ratios}
\label{sec-limit2d}

Following the Bayesian approach described above, a two-dimensional posterior p.d.f.,
       \begin{equation}
        \centering
f({\rm Br}_{u\gamma},{\rm Br}_{uZ}|\data),
       \end{equation}
was evaluated combining the HERA \Ronum{1} and HERA \Ronum{2} datasets.
Such a p.d.f. was built using the parameters described in \sect{theo} (no higher-order
QCD corrections were applied in this case) to
express the FCNC cross-section in terms of the anomalous top branching ratios.
The signal efficiencies for the different production channels ($\gamma$- or 
$Z$-mediated) and decay modes ($bW$ or $uZ$) were taken into account
(see \Tab{eff}). The selection efficiency of the $e$-channel is larger
for the $Z$-mediated process than the $\gamma$-mediated process, since in
this case the final-state electron is scattered at a larger angle and is 
more often visible in the detector.

The decay channel $t \rightarrow u\gamma$ was not simulated since the 
branching ratio is very low for the range of couplings under consideration. 
In addition, the selection efficiency is expected to be low for such events
 and was therefore set to zero.

\begin{figure}[p]
\vfill
\begin{center}
\includegraphics[scale=0.65]{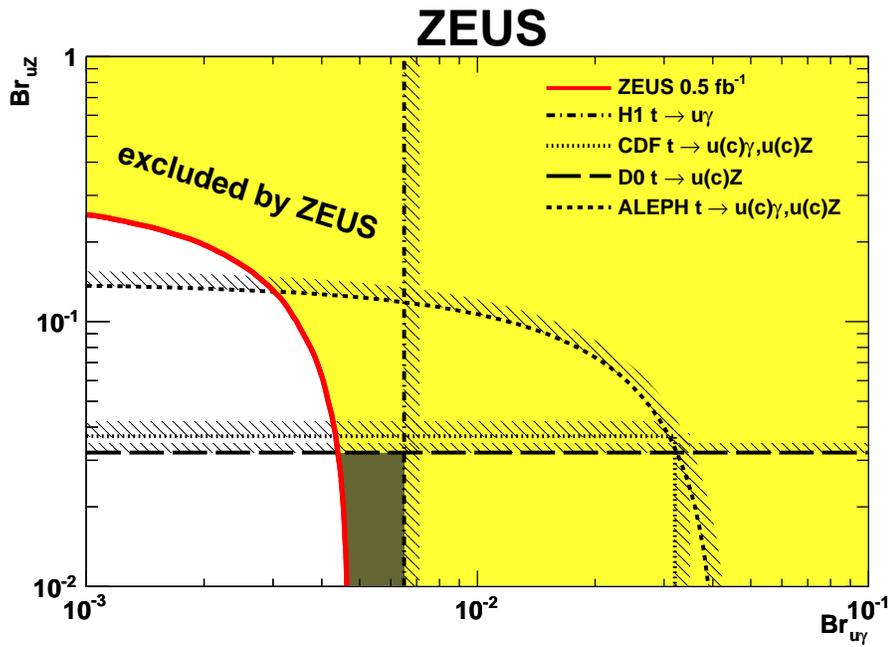}
\end{center}
\caption{ZEUS boundary in the (${\rm Br}_{u\gamma}$, ${\rm Br}_{uZ}$) plane.
Also shown are boundaries of {\rm H1~\protect\cite{singletoph1}}, 
{\rm CDF~\protect\cite{prl:80:2525}},
{\rm D0~\protect\cite{pl:b701:313}} and
{\rm ALEPH~\protect\cite{pl:b543:173}}.
The shaded area is excluded. The dark shaded region denotes the area uniquely excluded by ZEUS.}
\label{fig-brlim}
\vfill
\end{figure}

The $95\%~{\rm C.L.}$ boundary in the $({\rm Br}_{u\gamma},{\rm Br}_{uZ})$ plane
 was evaluated as the set of points
\begin{equation*}
\centering
f({\rm Br}_{u\gamma},{\rm Br}_{uZ}|\data)=\rho_0,
\end{equation*}
where $\rho_0$ was chosen such that

\begin{equation}
\centering
{\int\int}_{f({\rm Br}_{u\gamma},{\rm Br}_{uZ}|\data)>\rho_0} d{\rm Br}_{u\gamma} d{\rm Br}_{uZ} \ f({\rm Br}_{u\gamma},{\rm Br}_{uZ}|\data) = 0.95.
\end{equation}
\Fig{brlim} shows the ZEUS boundary in the (${\rm Br}_{u\gamma},{\rm Br}_{uZ}$)
plane compared to limits from H1~\cite{singletoph1} and from experiments 
at other colliders:
ALEPH~\cite{pl:b543:173} at LEP (other LEP experiments~\cite{pl:b521:181}
have similar results), CDF~\cite{prl:80:2525} and D0~\cite{pl:b701:313} 
at Tevatron. The $e^+e^-$ and
hadron colliders, contrary to HERA, have similar sensitivity to $u$- and 
$c$-quark;
their limits are hence on both decays $t\rightarrow qV$ with $q=u,c$.
The limits set by the ZEUS experiment in the region where ${\rm Br}_{uZ}$ is less than 
$4\%$ are the best to date. 


\section{Conclusions}
\label{sec-conclusions}
      A search for possible deviations from the Standard Model predictions due to 
      flavour-changing neutral current top production in events with 
      high-$p_T$ leptons and high missing transverse momentum was performed using an integrated 
      luminosity of $0.37 \fbi$,  collected by the ZEUS detector in 2004--2007.
      Since no significant deviation from the expectation was observed, 
      the results were used to put limits on the anomalous production of
      single top quarks at HERA.
    
      A  $95\%$ credibility-level upper limit on the cross section of $\sigma < 0.24 \pb$ at 
      a centre-of-mass energy of $318 \gev$ was obtained.
      The limit was combined with a previous ZEUS result, obtained using 
      HERA \Ronum{1} data, for a total integrated luminosity of $0.50 \fbi$, 
      giving a combined $95\%$ credibility-level upper limit of
      $\sigma <  0.13 \pb$ at $\sqrt{s}=315 \gev$. 
      This limit, assuming a vanishing coupling of the top quark to the $Z$ boson, $v_{Z}$, 
      corresponds to a constraint on the coupling of the top to the $\gamma$, $\kappa_{\gamma}$,
      of $\kappa_{\gamma} < 0.12$. Constraints on the anomalous top branching ratios 
      $t \rightarrow u\gamma$ and $t \rightarrow uZ$ were also evaluated assuming 
      a non-zero $v_{Z}$.
      For low values of $v_{Z}$, resulting in branching ratios 
      of $t \rightarrow uZ$ of less than $4\%$, this paper provides the current best limits.

\section*{Acknowledgements}
\label{sec-ack}

\Zacknowledge

\vfill\eject

{
\ifzeusbst
  \bibliographystyle{./BiBTeX/bst/l4z_default}
\fi
\ifzdrftbst
  \bibliographystyle{./BiBTeX/bst/l4z_draft}
\fi
\ifzbstepj
  \bibliographystyle{./BiBTeX/user/l4z_epj-ICB}
\fi
\ifzbstnp
  \bibliographystyle{./BiBTeX/bst/l4z_np}
\fi
\ifzbstpl
  \bibliographystyle{./BiBTeX/bst/l4z_pl}
\fi
{\raggedright
\bibliography{./BiBTeX/user/syn.bib,%
              ./BiBTeX/bib/l4z_zeus.bib,%
              ./BiBTeX/bib/l4z_h1.bib,%
              ./BiBTeX/bib/l4z_articles.bib,%
              ./BiBTeX/bib/l4z_books.bib,%
              ./BiBTeX/bib/l4z_conferences.bib,%
              ./BiBTeX/bib/l4z_misc.bib,%
              ./BiBTeX/bib/l4z_preprints.bib,%
              ./BiBTeX/bib/l4z_replaced.bib,%
              ./BiBTeX/user/myref.bib,%
              ./BiBTeX/bib/l4z_temporary.bib}}
}
\vfill\eject


\begin{table}[!htbp]
 \centering
  \begin{tabular}{c r c c}
    parameter & \multicolumn{1}{c}{value} & $M_t$ syst.  & PDF syst. \\
   \hline
    $A_{\sigma}$ & $7.71 \pb$   & $\mp 7\%$       & $\pm 4\%$\\   
   \hline
    $B_{\sigma}$ & $0.296 \pb$  & $\mp 7\%$       & $\pm 6\%$\\   
   \hline
    $C_{\sigma}$ & $-0.016 \pb$ & $-$             & $-$ \\   
   \hline
    $A_{\Gamma}$ & $0.299 \gev$  & $\pm 1\%$       & $-$\\   
   \hline
    $B_{\Gamma}$ & $1.36 \gev$   & $\pm 4\%$       & $-$\\   
   \hline
    $C_{\Gamma}$ & $1.48 \gev$   & $\pm 3\%$       & $-$ \\   
   \hline
  \end{tabular}
 \caption{\emph{Parameters used to evaluate single-top production cross sections and decay
  widths for the different channels. The systematic effects due to the uncertainties on the
  top mass and the parton distribution functions are also reported. 
  }}
 \label{tab-pars}
\end{table}


\begin{table}[!htbp]
\centering
 \begin{tabular}{l c c c}
  \multicolumn{4}{c}{}\\
   \hline
     & $N^{\rm obs}$ & $N^{\rm pred}$ & $W [\%]$ \\
   \hline
     electron channel $e^{+}p$ & 0 & 1.7$\pm$0.4 & 53 $\pm$ 11 \\  
     muon channel $e^{+}p$ & 1 & 1.5$\pm$0.2 & \multicolumn{1}{l}{64 $\pm$ 9} \\  
   \hline
     electron channel $e^{-}p$ & 1 & 1.9$\pm$0.4 & 51 $\pm$ 11 \\  
     muon channel $e^{-}p$ & 2 & 1.5$\pm$0.3 & \multicolumn{1}{l}{63 $\pm$ 9} \\  
   \hline
     electron channel $ep$ & 1 & 3.6$\pm$0.6 & \multicolumn{1}{l}{52 $\pm$ 9} \\  
     muon channel $ep$ & 3 & 3.0$\pm$0.4 & \multicolumn{1}{l}{64 $\pm$ 7} \\  
   \hline
 \end{tabular}
 \caption{\emph{Number of events passing the final selection cuts, $N^{\rm obs}$, compared to the
 SM prediction, $N^{\rm pred}$. The last column shows the $W$ contribution as a
      percentage of the total SM expectation. The uncertainties have been obtained by adding
      systematic and statistical contributions in quadrature. 
}}
 \label{tab-finalevts}
\end{table}


\begin{table}[!htbp]
 \centering
  \begin{tabular}{c l l c c c} \hline
  coupling &  \multicolumn{1}{c}{decay} & \multicolumn{2}{c}{$e-$channel} & \multicolumn{2}{c}{$\mu-$channel} \\ 
 & & \multicolumn{1}{c}{$\epsilon$} & $\Delta \epsilon/ \epsilon$ & $\epsilon$ & $\Delta \epsilon/ \epsilon$ \\ \hline
 $\kappa_{\gamma}$ & $t \rightarrow bW$ & $0.029$ & $\pm 0.04$ & $0.029$ & $\pm 0.04$\\ \hline
 $\kappa_{\gamma}$ & $t \rightarrow uZ$ & $0.0080$ & $\pm 0.08$ & $0.011$ & $\pm 0.07$\\ \hline
 $v_{Z}$ & $t \rightarrow bW$ & $0.048$ & $\pm 0.04$ & $0.024$ & $\pm 0.06$\\ \hline
 $v_{Z}$ & $t \rightarrow uZ$ & $0.066$ & $\pm 0.03$ & $0.012$ & $\pm 0.07$\\ \hline
\end{tabular}
\caption{\emph{Summary of selection efficiencies on signal samples for different production
couplings and decay modes. The relative errors are due to the statistics of the MC
samples.}}
\label{tab-eff}
\end{table}


\vfill


%
%
\end{document}